\documentclass[aps,floats,amsfonts,twocolumn]{revtex4}

\begin{document}

\newcommand{\beq}{\begin{equation}}
\newcommand{\eeq}{\end{equation}}
\newcommand{\bea}{\begin{eqnarray}}
\newcommand{\eea}{\end{eqnarray}}
\newcommand{\ba}{\begin{array}}
\newcommand{\ea}{\end{array}}
\newcommand{\om}{(\omega )}
\newcommand{\bef}{\begin{figure}}
\newcommand{\eef}{\end{figure}}
\newcommand{\leg}[1]{\caption{\protect\rm{\protect\footnotesize{#1}}}}

\newcommand{\ew}[1]{\langle{#1}\rangle}
\newcommand{\be}[1]{\mid\!{#1}\!\mid}
\newcommand{\no}{\nonumber}
\newcommand{\etal}{{\em et~al }}
\newcommand{\geff}{g_{\mbox{\it{\scriptsize{eff}}}}}
\newcommand{\da}[1]{{#1}^\dagger}
\newcommand{\cf}{{\it cf.\/}\ }
\newcommand{\ie}{{\it i.e.\/}\ }
\newcommand{\eg}{{\it e.g.\/}\ }

\title{Contextual objectivity and the quantum formalism.}
\author{Philippe Grangier}
\address{Laboratoire Charles Fabry de l'Institut d'Optique, %\\
F-91403 Orsay, France}

\begin{abstract}

The ``new orthodoxy" of quantum mechanics (QM) based on the decoherence 
approach  requires many-worlds  as an essential ingredient  
for logical consistency, and one may wonder what status to give to
all these ``other worlds". 
Here we advocate that it is possible to build a consistent approach to QM 
where no other worlds are needed,
and where the quantum formalism appears as a consequence
of requiring the enumerability of physical properties. 
Such a quantization hypothesis is closely related to indistinguishability,
and is deeply inconsistent with classical physics. 

\end{abstract}

\maketitle

\section{A short status report.} 

The present orthodoxy of quantum mechanics - and even of physics in 
general - takes for granted that since the ultimate physical theory must 
a quantum theory, then it must be possible to reconstruct classical 
physics as a suitable approximation of quantum physics. For instance, 
classical physics should emerge from a suitable coarse-grained 
average done on localized objects, which is what we have at hand after 
a quantum system has undergone a decoherence mechanism. 

Though the attempts done in that direction may be partially convincing, 
they hit what we consider to be an important conceptual difficulty, which 
is that any ``decohered" quantum system sits in a ``multiverse", or 
``branching universe", where the different possible outcomes of a 
quantum measurement (e.g. a Stern-Gerlach experiment) have exactly 
the same status. In order to interpret the trivial fact that only one result 
is found in a given measurement, and may decide about the whole 
future of our ``unique" universe, one has to assume that we ``live and 
perceive in one branch only", and moreover that no further 
communication is possible between the various branches of the 
multiverse. Though one may  consider this set of ideas as 
a definition of randomness, we cannot prevent ourselves to find it 
deeply unsatisfactory. Actually, it appears that such an attempt to give 
an ontological status to a quantum state leads to a complete 
mess-up of the very idea of ontology, since ``all possibles" 
acquire exactly the same ontological status. 

As a back-up, an alternative interpretation is the ``old orthodoxy" provided by 
the Copenhagen interpretation of QM. Though this is really old-fashioned indeed,
it has a modern version, that we can call the ``minimalistic" interpretation of QM~:
it consists of looking at quantum states as algorithms for computing correlations
between successive measurement outcomes. This view is quite nice and useful - 
and in some sense more solid than the previous ones - but it has the big 
disadvantage that it makes it very difficult to ask any question like~:
Why is QM the way it is~? 
Such a question, having nothing to do with correlations between 
successive measurement outcomes, is simply heretic in this context, and thus cannot
be answered. However, QM is now almost one century old, and the questions about
how the baby was born should not be censored any longer. 

\section{An alternative approach.} 

A leading  line of our alternative approach will be  the following~: since 
we know that the Copenhagen interpretation ripped to the bones - that 
is,  the minimalistic interpretation -  has always been  quite 
successful, is there a way to see where it comes from - or in some 
sense, to make it necessary ? 

Actually, the Copenhagen interpretation has a built-in feature, which is 
the infamous ``collapse of the wave function". This collapse was reborn 
under a more civilized form in the minimalistic interpretation. which 
accepts only questions about correlations, and thus  does not need to 
speak about any collapse. But one essential feature remains~: a 
quantum measurement gives only one result, not many ones, in sharp 
opposition with the multiverse approach quoted above. 

Thefore our line of attack will be the following~: we will keep as 
close as possible to the  minimalistic interpretation, and we will make 
more precise what actually occurs when a measurement is done~: 
under idealized conditions, this simply defines the (quantum) state of the 
system, as explained in \cite{ph1,ph2,ph3,ph4}.
As in the minimalistic interpretation, 
there is no need for adding any  further collapse~: all is already 
included in the fact  that measurements define a state - and thus 
other measurements will simply define another state. 
Following this approach,
 a quantum state will be generally defined as the values of a set of physical 
properties, which can be predicted with certainty and measured
reproducibly without changing the system\cite{ph1}. 
The  set of physical  properties associated with all measurements
needed to define a state
constitutes a ``context" (using the terminology
of standard QM, this is a ``complete set of commuting observables"). 
In a given context, all possible states constitute a set of
``exclusive modalities", that is, 
they are associated with measurement outcomes which can be
unambiguously identified.  This allows us to 
give an objective status to the quantum state, which 
is associated with predictions that are certain 
and observer-independant in a given context, 
hence the wording ``contextual objectivity". 

Now the question is~: why is this state a quantum state, and not a 
classical one ? The intuitive answer we want to give is~: ``the state 
is quantum when the exclusive modalities
are quantized, while the parameters defining the contexts are not". To 
make it clearer through an example, 
throwing a dice has a quantized outcome (a number 
between one and six)  but also  a quantized number of measurable 
physical quantities, because it is assumed that the dice always lies on 
one side. But if it is assumed that the dice can be oriented in space along 
any angles (using e.g. Euler angles), but that the measurement still  
gives only six outcomes whatever the dice orientation  - then it 
becomes a quantum dice. 

We may thus give the following reasonable postulate~:  a quantum 
system is characterized by the fact that any complete set of 
measurements will always provide a fixed number N of possible 
exclusive modalities. The measurements themselves may depend on continuous 
parameters, so the number of possible measurements is in principle 
infinite.  Hovever,  the number of mutually exclusive 
possible answers should be only N. 

As a simple example, let us consider the case where a quantum measurement 
depending on one continuous angle $\alpha$ gives two results $+$ or 
$-$. From our postulate, the same  measurement  done for the angle 
$\beta \neq \alpha$ will also give two results $+$ or $-$. Classically the results for 
$\beta$ can be considered as ``refinement" of the properties obtained 
for $\alpha$, and one should be able to define four exclusive modalities,
$(\alpha:+,\;\beta:+)$, $(\alpha:+,\;\beta:-)$ , $(\alpha:-,\;\beta:+)$,
$(\alpha:-,\;\beta:-)$. But this contradicts our hypothesis (and also
quantum mechanics)~: we have assumed 
that the number of exclusive modalities is {\bf only two}, and therefore 
it cannot be split further by performing more measurements. 

But why postulate that the number of exclusive measurement 
outcomes should be fixed ? Actually, such a quantization postulate 
seems to be required to give a meaning to the fact that  two systems 
are ``identical"~: two systems will  be identical only if is possible to 
enumerate all their properties, and if they are all the same. It was very 
convincingly recognized by Leibnitz that identity cannot exist in classical 
physics, because one smaller detail is always able to distinguish two 
objects. He concluded that identity (or indistinguishability) is thus 
impossible, but here he was wrong~: identity {\bf is} possible, but quantum 
physics is required to give it a meaning.

\section{Reconstructing the quantum formalism ?} 

Now, is the fact that N is fixed enough to deduce the whole formalism of 
quantum mechanics ? Though we cannot prove it fully presently, we 
are very tempted to answer positively. The main lines of the reasoning 
are the following~:

(i) we consider two different contexts, that is two different complete 
sets of measurements performed on the system. For any such 
context (there is a multiple continuous infinity of them),  there are only 
N exclusive measurements outcomes, that will be called ``exclusive 
modalities" (orthogonal quantum states in the usual QM formalism). 

(ii) since we postulated that the N exclusive modalities in one context 
cannot  be ``put together" with the N exclusive modalities in another 
context, the modalities taken from two different contexts are essentially 
``non-exclusive". In other words, it is impossible to ``give more details" 
so that non-exclusive modalities would become exclusive, because we 
would end up in having more than N exclusive modalities. 

Therefore, 
the only meaningful question we can ask to the theory is~: if the system 
is in modality $a_i$ of context $E$, what is the {\bf probability} $p_{j|i}$
to find it in modality $b_j$ of context $E'$ ? 
We emphasize that probabilities 
appear here to be logically needed for consistency~: they are actually 
the only way to keep the hypothesis of having only N exclusive 
modalities consistent with the infinite number of contexts. In some 
sense, this is the trade-off~: we don't want many worlds, but we 
absolutely need probabilities. 

Then another question arises : is there any relation between 
the above $p_{j|i}$, and the reciprocal probability $p_{i|j}$
to find the system in modality $(a_i, E)$, knowing that it 
is in modality $(b_j, E')$~?
(it should be clear that in the framework of standard QM,
one has $p_{i|j} = p_{j|i}$).
Here we may consider  that 
we are actually looking for a joint probability  $p_{ij}$ 
connecting the 
modalities $(a_i, E)$ and $(b_j, E')$. However,
the notion of a joint probability takes a special meaning,
because the $N^2$ ``events"
$(a_i,E \text{ and } b_j,E')$ {\bf do not} correspond
to $N^2$ exclusive modalities for the system (this is forbidden 
by our main hypothesis). %Therefore, there is no
%situation where they might be predicted with probability one. 
Nevertheless, in order to keep as close as possible 
to the notion of a symmetrical joint probability, we will 
simply assume that $p_{ij} = p_{i|j} = p_{j|i}$.
Then  $p_{ij}$ depends symmetrically
on the two contexts $E$ and $E'$,  and correspond to the probability
for the system to be in ``both" modalities $(a_i,E)$ and $(b_j,E')$. 
This (non-classical) ``both" is rigourously defined from 
the conditional probabilities, which keep a clear
operational meaning in our case.

(iii) therefore, the theory must provide a $N \times N$ probability matrix 
$\Pi = \{ p_{ij} \}$, giving the probabilities  $p_{ij}$ 
connecting the  modalities $(a_i, E)$ and $(b_j, E')$
as defined above. 
It is then simple to check that $\Pi$ is a
doubly stochastic matrix (all lines and columns sum to one), which 
reduces to the identity if $E=E'$. 

In order to manipulate these matrices, 
it is convenient to introduce orthogonal $N \times N$ projectors,
by attributing a set of orthogonal projectors $\{ \pi_i \}$ 
(resp. $\{ \pi_j' \}$) 
to the exclusive modalities $\{ a_i \}$ (resp. $\{ b_j \}$) of 
the context $E$ (resp. $E'$). 
Then the question arises whether is 
possible to define an arbitrary doubly stochastic matrix as a function of 
these projectors. The answer is yes, by simply taking~: $p_{ij} = 
Trace(\pi_i \pi_j')$. This requires however that complex numbers 
(i.e. hermitian projectors)
are used, because it is not possible to build an arbitrary doubly 
stochastic matrix from  two sets of real projectors
(there are not enough variables). Our projectors
are therefore hermitian $NÊ\times N$ matrices. 

Now we are almost done, since a set of $N$ hermitian projectors $\{ 
\pi_i \}$ can be associated with a set of rays in the 
corresponding Hilbert space, that is with usual quantum states. 
Moreover, we already know that in context $E$ the measurement of the 
physical quantity $A$ will give with certainty the result
$\alpha_i$ in state $\pi_i$. This is consistent with associating
to A the operator $\hat A = \sum_i \alpha_i \pi_i$, and thus 
recovering the standard definition of observables.

Another crucial feature is that the 
change between context $E$ and context $E'$
appears to  be associated with a unitary matrix $\Sigma$, so that 
$ \pi_j'  = \Sigma^\dagger  \pi_i \Sigma$. 
Since these matrices must also act on the operators $\hat A$,
one should require that they provide a {\bf representation}
(in the sense of groups theory) of the geometrical transformations
acting of the true physical quantities. 
Depending on the geometrical group which is considered
(rotation group, Galileo group, Poincar\'e group), one
should thus be able to reconstruct various kinds of 
Hilbert spaces (angular momentum, non-relativistic QM, relativistic QM...).

It is also important to note  that the matrices 
$\Sigma$ connecting different contexts should not commute, 
otherwise one would come back simply to a classical 
alternative between N exclusive modalities. 
As said before, having a discretized 
number of exclusive modalities is not enough to get quantum mechanics,
one needs also that the measured physical quantities (the contexts) are connected
by non-commutative geometrical transformations.

Finally, an interesting question is where $\hbar$ comes in.
This can be illustrated by taking the example of angular momentum,
associated with 3-D rotations
acting on the true physical quantities.
Following our approach, we have to build a unitary representation
of the rotation group, which is well known to be given by
$U_{{\bf u}} = \exp(-i \; {\bf u}.{\bf \hat j})$.
Here ${\bf u}$ is a 3-D vector defining a rotation $R_{{\bf u}} $,
and ${\bf \hat j}$ is a set of 3 (dimensionless) $N \times N$ matrices
$\hat j_{x,y,z}$,
obeying the angular momentum commutation relations obtained
from the Lie algebra of the rotation group. 
By a standard calculation we can get the eigenvalues and eigenvectors
of the complete set $\{ {\bf \hat j}^2, \; \hat j_z \}$. 
Now, one can also physically measure the angular momentum of the system,
by using a Stern and Gerlach apparatus, and thus construct the
``physical" observable ${\bf \hat J}$, where $\{ {\bf \hat J}^2, \; \hat J_z \}$
can be associated to a given context. It appears then that ${\bf \hat j}$
and ${\bf \hat J}$ have exactly the same behaviour, up to a multiplicative 
constant which is precisely $\hbar$. 

The rotation case is especially simple because $\hbar$
has the dimension of an angular momentum. 
But if one takes for instance translations, the rule is just the same~:
the unitary transformation is $T_{{\bf a}} = \exp(-i \; {\bf a}. {\bf \hat p})$
where ${\bf a}$ has dimension $L$ (length) and ${\bf \hat p}$ has thus dimension $1/L$. 
Then the observable ${\bf \hat P}$ constructed from the measurements results
will fit with ${\bf \hat p}$ simply by writing ${\bf \hat P} = \hbar \; {\bf \hat p}$. 
This reasoning clearly shows that $\hbar$ does appear
as the ``unit" of quantization, which is required to connect the two
(``probabilistic" and ``geometrical") definitions 
of an observable which appear in our construction. 

Therefore, the dimension of $\hbar$ is an action because it has to match
a physical observable $\hat A = \sum_i \alpha_i \pi_i$ onto
an infinitesimal generator, which has the dimension
of the inverse of the physical quantity conjugated to $\hat A$. 
We note that this provides a view of canonically conjugated observables
directly in the framework of group theory.

\section{Conclusion}

Though many assertions are still to be proven, our program  is to establish 
that the quantum formalism is a consequence of the quantization 
hypothesis, and that this formalism is actually the (only ?) way to make 
quantization consistent. 
 
The quantum ``strangeness" is thus rooted in the fact that 
``enumerability" of properties, which is closely related to indistinguishability,
is deeply 
inconsistent with classical physics. We emphasize again that our ``fixed N" 
postulate should not be confused with classical discreteness, which is 
an approximation always subject to further refinement. The fact that 
the ``gain of knowledge" must stop  because there is ``nothing" 
between zero and one is actually the crucial quantum hypothesis. 
 
As a conclusion, here are a few more remarks~:
 
- in our approach a  quantum measurement is the operation by which we 
define a quantum state, and there is essentially no difference between 
the beginning (``preparation of the state") and the end (``collapse of 
the state"). Therefore the so-called collapse problem essentially vanishes.
In case you feel unhappy with this, a convenient retreat is the minimalistic
interpretation introduced above~: QM provides a way to compute correlations between 
successive measurement results. 
 
- similarly, there is no need for many worlds. 
A quantum state is always defined relative 
to a classical context, in which the state-defining measurement is 
carried out. If the new context is incompatible with the previous ones 
(that is, if the new modalities are non-exclusive with the previous ones), 
the state is reset as one among the new exclusive modalities. This 
does not result from any mysterious ``branching of universes", but it is the 
unavoidable consequence of the quantum postulate, and the only 
way to keep the enumerability of exclusive modalities  consistent with 
the continous infinity of contexts. 
 
- the usual question of the ``boundary" between the classical and the quantum 
world is ill posed. Presently, quantum mechanics is always formulated in 
a classical context, where Leibnitz's intuition is correct~: properties are 
not enumerable, and objects are never identical. But at the quantum 
level Leibnitz is wrong, and just the opposite becomes true. This 
difference  makes a quite decent boundary, which is enough for all 
practical purposes. 
%To go further, one should look at questions such 
%as ``is it possible to formulate quantum mechanics from inside~?",
%which are also probably ill posed. 

- our approach may also suggest new questions~: 
is it possible to formulate QM without any reference to classical physics ? 
or to build  a ``time-ordered" theory where $p_{i|j} \neq p_{j|i}$ (see above)~? 
can we give a simple geometrical meaning
to all physical quantities in a unit system where $\hbar = 1$~?  
Further considerations about entanglement, Bell's inequalities, 
and related subjects
can be found in previous eprints and publications \cite{ph1,ph2,ph3,ph4,ph5}.

{\bf Acknowledgements.}
Many thanks are due to the organizers and participants of the
meeting ``Foundations of Quantum Information" (April 2004, Camerino, Italy), 
for many stimulating discussions about the topics presented in this paper, 
and to Franck Laloe for continued support.

\end{document}